\documentclass{ws-jai}

\title{A GPU Spatial Processing System for CHIME}
\author{
Nolan Denman$^{123*}$, 
Andre Renard$^{1}$, 
Keith Vanderlinde$^{12}$, 
Philippe Berger$^{45}$,\\ 
Kiyoshi Masui$^{6}$, 
Ian Tretyakov$^{17}$, 
and the CHIME Collaboration
	}

\address{
$^{1}$Dunlap Institute for Astronomy and Astrophysics, University of Toronto, Toronto, Ontario, M5S 3H4, Canada\\
$^{2}$Department of Astronomy and Astrophysics, University of Toronto, Toronto, Ontario, M5S 3H4, Canada\\
$^{3}$Central Development Laboratory, National Radio Astronomy Observatory, Virginia, 22903, USA\\
$^{4}$Canadian Institute for Theoretical Astrophysics, University of Toronto, Toronto, Ontario, M5S 3H4, Canada\\
$^{5}$Jet Propulsion Laboratory, California Institute of Technology, Pasadena, California, 91109, USA\\
$^{6}$MIT Kavli Institute for Astrophysics and Space Research, Massachusetts Institute of Technology, Cambridge, Massachusetts, 02109 USA\\
$^{7}$Department of Physics, University of Toronto, Toronto, Ontario, M5S 3H4, Canada
}

\usepackage{natbib}
\usepackage[flushleft]{threeparttable}

\usepackage{hyperref}

\usepackage{graphicx}
\graphicspath{{./images/}}

\usepackage{amsmath}
\usepackage{amssymb}
\usepackage{aas_macros}
\usepackage{microtype}
\usepackage{lmodern}

\mathchardef\myphen="2D

\newcommand{\ndat}{\hspace*{\fill}---\hspace*{\fill}}

\usepackage{booktabs}

\usepackage{multirow}

\begin{document}

\maketitle

\begin{history}
\received{(to be inserted by publisher)};
\revised{(to be inserted by publisher)};
\accepted{(to be inserted by publisher)};
\end{history}

\begin{abstract}
We present an overview of the Graphics Processing Unit (GPU) based
spatial processing system
created for the Canadian Hydrogen Intensity Mapping Experiment (CHIME).
The design employs AMD S9300x2 GPUs
and readily-available commercial hardware
in its processing nodes
to provide a cost- and power-efficient 
processing substrate.
These nodes are supported by a liquid-cooling system
which allows continuous operation
with modest power consumption and 
in all but the most adverse conditions.
Capable of continuously correlating 2048 receiver-polarizations across 400\,MHz of bandwidth,
the CHIME X-engine constitutes
the most powerful radio correlator currently in existence.
It receives $6.6$\,Tb/s of channelized data from CHIME's FPGA-based F-engine,
and the primary correlation task
requires  $8.39\times10^{14}$ complex multiply-and-accumulate operations per second.
The same system also provides 
formed-beam data products 
to commensal FRB and Pulsar experiments;
it constitutes
a general spatial-processing system
of unprecedented scale and capability,
with correspondingly great challenges
in computation, data transport, heat dissipation, and interference shielding.
\end{abstract}


\corres{$^*$Corresponding author, \url{ndenman@nrao.edu}}

\section{Introduction}
\label{sec:x-eng-intro}

The Canadian Hydrogen Intensity Mapping Experiment
(CHIME, \url{https://chime-experiment.ca})
is a purpose-built instrument located at the 
Dominion Radio Astrophysical Observatory (DRAO)
near Penticton, British Columbia, Canada.
It consists of four adjacent 100\,m long and 20\,m wide semi-parabolic cylinders, 
each instrumented with 256 dual-polarization receivers 
optimized for observations between 400 and 800\,MHz.
It is designed to constrain the physical nature of Dark Energy
by observing its effects on the geometry of the universe
\citep{2013PhR...530...87W};
it will accomplish this by using the Baryon Acoustic Oscillation feature
in the large-scale distribution of matter as a ``statistical standard ruler'',
observed through the 21\,cm emission of neutral Hydrogen 
\citep{2008PhRvL.100p1301L,2012RPPh...75h6901P}
at cosmological redshifts of 0.8 to 2.5.
Additional scientific programs focused on Pulsar observation \citep{CHIMEPSR:inprep, 2017arXiv171102104N}
and Fast Radio Burst (FRB) detection \citep{2018ApJ...863...48C}
make commensal use of the wealth of astronomical data obtained.

As an interferometric radio telescope,
CHIME requires a spatial processing system
capable of combining the signals from its receivers 
to produce a view of the sky \citep{2009tra..book.....W}.
In order to form the astronomically-crucial `visibilities',
the time-domain electric field measurements from each receiver
must be Fourier-transformed
(or `channelized' -- separated into frequency components)
and the signal from each receiver correlated against every other receiver.
The specific parameters of CHIME,
particularly the large numbers of receiver-polarizations ($N=2048$) and frequency channels ($M=1024$),
make an `FX' correlator architecture
(in which the channelization precedes the inter-receiver product)
substantially more efficient than an `XF' architecture
(in which the product of receiver signals precedes the Fourier transform)
\citep{1999ASPC..180...57R,2001isra.book.....T}.
The computational cost of the former scales as $N\log M+N^2$
to the latter's $N^2M$,
and is therefore more efficient in the limit of many frequency channels.

Hybrid correlator systems
using Field-Programmable Gate Arrays (FPGAs) for the Fourier-transform-stage
and Graphics Processing Units (GPUs) for the outer-product-stage
have seen widespread adoption in recent correlator designs
(including several of those in Table~\ref{tab:corrs}).
FPGAs combine flexible signal-processing capabilities
with substantial data transport resources,
while
GPUs are purpose-built for large parallel matrix and vector operations,
well-suited to the demands of a correlator X-engine
\citep{2011arXiv1107.4264C}.
The CHIME correlator X-engine therefore provides
a fully-functional example of a
powerful and cost-effective
component of a hybrid correlator system.

\S\ref{ssec:over} provides an overview of the CHIME signal processing path.
The CHIME correlator X-engine
distributes processing tasks between
a large number of GPU-hosting `nodes',
whose design and composition
were determined by the intersection of the
primary data transport and processing requirements
described in \S\ref{ssec:cmput}
and the 
mass-market electronics
available at the time of their construction.
The design of the X-engine as a whole focuses on
housing the GPU nodes,
supplying them with electrical power,
keeping them thermally stable, and
handling their incoming and outgoing data.
\S\ref{ssec:node-hw} and \S\ref{ssec:infra}
describe in detail
the X-engine nodes and their supporting infrastructure,
respectively.
\S\ref{sec:discuss} illustrates the 
computational and data-transfer 
performance of the X-engine system (\S\ref{ssec:cmput-perf}),
the system's power consumption (\S\ref{ssec:power}),
and the success of the system's cooling infrastructure (\S\ref{ssec:cool-perf}).
Discussion follows in \S\ref{sec:conc},
with commentary regarding potential future developments in \S\ref{ssec:future}.

\section{CHIME Spatial Processing System Description}
\label{sec:full-x}

\subsection{CHIME Signal Processing System Overview}
\label{ssec:over}

\begin{figure}[ht]
    \centering
    \includegraphics[width=0.6\textwidth]{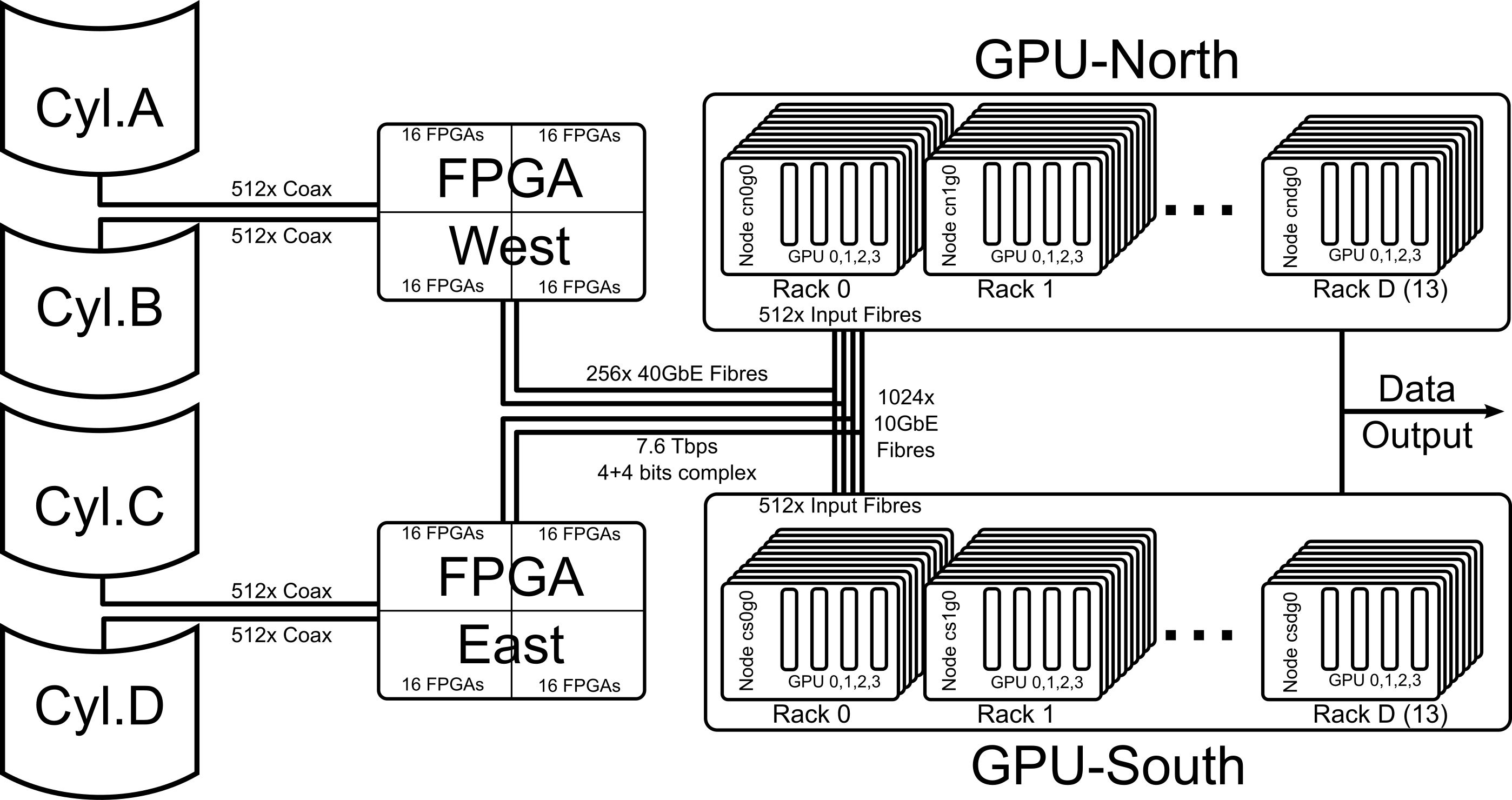}
    \caption{
    Schematic overview of data flow throughout the CHIME signal processing system,
    as described in \S\ref{ssec:over}.
    }
    \label{fig:system-schematic}
\end{figure}

\begin{figure}[ht]
    \centering
    \includegraphics[width=\textwidth]{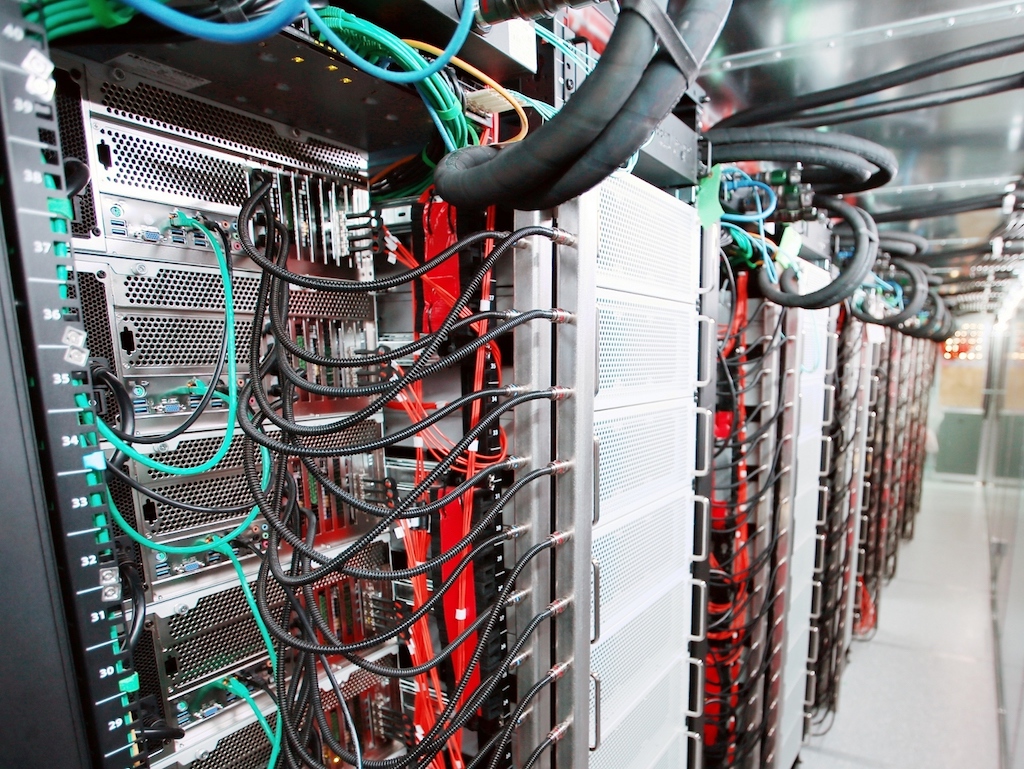}
    \caption{
    	One of the two CHIME GPU X-engine enclosures;
    	visible components include     	
        the GPU nodes,
        secondary coolant distribution manifolds,
        network connections to the F-engine (orange fibre)
        and the external network (green and black cables).
    }
    \label{fig:cluster}
\end{figure}

The overall structure of the
CHIME signal processing system
is shown in Figure~\ref{fig:system-schematic},
with key parameters collected in Table~\ref{tab:ch-stats}
and a photograph of the completed correlator
in Figure~\ref{fig:cluster};
further description follows.

\begin{wstable}[ht]
    \centering
    \begin{tabular}{lclc}\toprule
    Number of Receiver-Polarizations & 2048 & Observing Band & 400-800\,MHz \\
    Collecting Area & 8000\,m$^2$ & Number of Frequency Channels & 1024  \\
    Receiver Noise Temperature & $\approx50$\,K & Frequency Resolution & 390\,kHz \\
    North-South Field of View & $\sim120^\circ$ & 
    East-West Field of View & $2.5^\circ$-$1.3^\circ$ \\
    \bottomrule
    \end{tabular}
    \caption{
    Selected observational parameters of the CHIME instrument.
    }
    \label{tab:ch-stats}
\end{wstable}

CHIME's light-gathering apparatus
consists of four semi-parabolic dishes,
each 100\,m in length and 20\,m in width
with a 5\,m focal length.
The central 80\,m of each cylinder is instrumented
with 256 custom-designed 
dual-polarization clover-leaf receivers \citep{6887670}
spaced $\approx 30.48$\,cm apart,
for a total of 2048 distinct receiver-polarization `inputs' to the correlator system.

The signal from each receiver-polarization is
separately amplified and transported 
along coaxial cable to its cylinder's
adjacent F-engine enclosure.
There, it is further filtered and amplified
before being fed to an FPGA-based 
digitizer and Fourier-transform system \citep{2016JAI.....541005B}.
This system consists of 128 custom-built motherboards,
which employ the Xilinx Kintex-7 420T FPGA\footnote{
\url{https://www.xilinx.com/products/silicon-devices/fpga/kintex-7.html}}
for signal processing.
A single motherboard hosts two daughter boards,
each of which provides 
the coaxial connectors for four receiver-polarizations of RF signal input.
Each daughter board also houses 
two Teledyne EV8AQ160 Analog-to-Digital Converters\footnote{
\url{https://www.teledyne-e2v.com/products/semiconductors/adc/ev8aq160}}
whose output is transferred to the motherboard
as input to a Polyphase Filter Bank and Fast Fourier Transform.

At this point, 
each FPGA has data from only one receiver-polarization
but over the instrument's full 400-800\,MHz frequency range;
an FX architecture requires this data be regrouped
(turned into sets of single-frequency-channel data from all receiver-polarizations)
to permit the construction of full-array visibilities.
This regrouping can be thought of as the transposition of 
a two-dimensional data array in receiver-frequency space,
and is often referred to as a `corner turn' in correlator-specific literature.
The backplane within each `crate' of 16 FPGAs 
and high-speed inter-crate connections
lack the capacity to complete the transposition of the data,
assembling the complete cylinders (but not the full array)
in the final inter-crate transfer stage
(see \citet{2016JAI.....541004B} for details).
The data is therefore transferred to the X-engine 
as `bundles' of four frequency channels, 
each of which has data from only one cylinder.
Four of these bundles are combined in each of the X-engine's GPU nodes
to provide the required full-array coverage for all four frequencies.

CHIME's spatial processing
takes place on 256 custom-built GPU nodes, 
each processing four frequency channels,
which collectively perform the full outer-product correlation of 2048 inputs
for each of the 1024 frequency channels
every 2.56 microseconds.
These nodes additionally form a set of 10 phased-array beams
for pulsar timing \citep{CHIMEPSR:inprep, 2017arXiv171102104N}
and perform up-channelization and Fourier-transform-based beam formation
on an additional copy of the data 
prior to its export to the FRB back-end \citep{2017arXiv170204728N}.
Additional parallel processing streams are currently under development.
The X-engine GPU nodes reside in a pair of purpose-built enclosures,
and are supported by a multi-stage liquid-based cooling system;
these are further described in \S\ref{ssec:infra}.

Post-correlation, visibility data 
is averaged and recorded for later physical transport
to Compute Canada facilities
for processing and analysis.
Beams formed for the Pulsar and FRB components of the system
are exported over the site's internal network for further processing.

\subsection{Computational Requirements}
\label{ssec:cmput}

For an FX-architected correlator, 
the cost of the inter-receiver product 
dominates the overall computational requirements of visibility formation.
For an instrument with $N$ receiver-polarizations and instantaneous bandwidth of $\Delta\nu$
the na\"{\i}ve computational cost of the `X' stage -- including autocorrelations -- is
$\eta = \Delta\nu\cdot\frac{N(N+1)}{2}$ complex multiply-accumulate (cMAC) operations per second.
CHIME therefore requires $8.39\times10^{14}$\,cMAC/s $\approx$\,0.8\,PcMAC/s 
to complete this stage of its correlation.
For comparison, the computational requirements (as measured by $\eta$)
of other radio interferometers are shown in Table~\ref{tab:corrs}.

\begin{wstable}[ht]
    \centering
    \begin{tabular}{llccc}\toprule
    Instrument & Reference & $N$ & $\Delta\nu$ (MHz) & $\eta$ (TcMAC/s)\\\midrule
    CHIME$^\dagger$  & \textit{hic} & 2048 & 400 & 839 \\
    SKA LFAA & \citet{SKAMEMO} & 1024 & 300 & 157 \\
    HERA--350$^\dagger$ & \citet{2017PASP..129d5001D} & 700 & 200 & 49 \\
    MWA Phase II$^\dagger$ & \citet{2018arXiv180906466W} & 512 & 327 & 43 \\
    CHIME Pathfinder$^\dagger$ & \citet{2014SPIE.9145E..22B} & 256 & 400 & 13.6 \\
    MWA Phase I$^\dagger$ & \citet{2015PASA...32....6O} & 256 & 327 & 10.8 \\
    OVRO LEDA$^\dagger$ &\citet{2015JAI.....450003K} & 512 & 58 & 7.6 \\
   	MeerKAT Phase 1 & \citet{2012AfrSk..16..101B} & 128  & 750 & 5.9\\   	
    ALMA ACA & \citet{ALMASPEX} & 64 & 2000 & 4.2 \\
    PAPER--128$^\dagger$ & \citet{2015ApJ...809...61A} & 256 & 100 & 3.3\\
    \midrule
    ngVLA (proposed)$^*$  & \citet{2018SPIE10700E..1OS} & 526 & 20000 & \hspace{6pt}2777\\
    ALMA BLC$^*$ & \citet{ALMASPEX} & 64 & 16000 & \hspace{6pt}33 \\ 
    EVLA WIDAR$^*$ & \citet{2009IEEEP..97.1448P} & 27 & 16000 & \hspace{6pt}6.0 \\
    \bottomrule
    \end{tabular}
    \caption{
    The computational requirements of other radio X-engines,
    as measured by the $\mathcal{O}(N^2\Delta\nu)$ metric $\eta$.
    Several of these ($\dag$) are FX hybrid FPGA-GPU designs
    as described in \S\ref{sec:x-eng-intro}.
    The ALMA BLC, EVLA WIDAR, and ngVLA correlators ($*$)
    use or plan to use `hybrid-XF' or `FFX' architectures,
    making direct comparisons difficult.
    }
    \label{tab:corrs}
\end{wstable}

The CHIME X-engine
was originally envisioned as a Fourier-transform-based system \citep{2009PhRvD..79h3530T, 2010PhRvD..82j3501T}
which exploited redundancies in baseline geometry to permit efficient $\mathcal{O}(N{\log}N)$ spatial correlation.
Interest in alternative correlation techniques
as well as the requirements of the CHIME/Pulsar and CHIME/FRB experiments
placed a premium on the ability to modify the X-engine software
quickly and with minimal development effort.
The goal of an adaptable, extensible, and re-configurable
correlator system
motivated 
the selection of GPUs as a both powerful and flexible
processing substrate.
This flexibility, along with the increasing computational power and efficiency of mass-market GPUs,
allowed a variety of algorithmic improvements and optimizations
which ultimately enabled a full-correlation $N=2048$ X-engine
which runs at a 100\% duty cycle
in parallel with a variety of 
beam-forming and spectral up-sampling tasks.

\begin{wstable}[ht]
    \centering
    \begin{tabular}{lclc}\toprule
    Number of Nodes & 256 & Number of Enclosures & 2\\
    Number of Racks & 26 & Nodes per Rack & 8 or 10 \\
    CPUs per Node & 1 & GPUs per Node & 2 dual-chip \\
    Total Main Memory & 32\,TiB & Main Memory per Node & 128\,GiB\\
    \bottomrule
    \end{tabular}
    \caption{
    Select parameters of the CHIME X-engine as a whole;
    see also Tables~\ref{tab:data-flow} and \ref{tab:node-pwr}.
    }
    \label{tab:sys-stats}
\end{wstable}

\begin{wstable}[ht]
    \centering
    \begin{tabular}{lcc}\toprule
    Parameter & per Node & per GPU \\
    \midrule
    Spectral Channels & 4 & 1 \\
    Observing Bandwidth& 1.56\,MHz & 390\,kHz\\
    Required cMAC/s & $3.3\times10^{12}$ & $8.4\times10^{11}$ \\
    \bottomrule
    \end{tabular}
    \caption{
    Select parameters of the CHIME X-engine nodes;
    see also Tables~\ref{tab:data-flow} and \ref{tab:node-pwr}.
    }
    \label{tab:node-stats}
\end{wstable}

The total data transfer bandwidth required between the F- and X-stages is
$\approx N\Delta\nu d$ for data with a total bit depth of $d$. 
In the case of CHIME, $d$ is 8 bits (4 bits each for the real and imaginary components),
resulting in a minimum of $6.6$\,Tb/s.
Associated flags and metadata (644\,B for each 4096\,B of data) further increase this data volume,
resulting in a total input bandwidth of $7.6$\,Tb/s.\footnote{
As a comparison, $7.6$\,Tb/s is 2490 PB/month; 
total global IP traffic in 2019 was estimated at 201,000 PB/month \citep{cisco-ip}.
CHIME therefore requires F--X data transfer at a rate equal to $1.2\%$ of global Internet bandwidth.}

\subsection{The GPU Processing Nodes}
\label{ssec:node-hw}

Post-Fourier-transform,
the data for each frequency channel may be processed independently;
this was therefore selected as the axis 
across which to distribute the processing tasks.
The fundamentals of the array 
dictate that correlating each $\approx390$\,kHz frequency channel requires 
$\approx7.4$\,Gb/s of input bandwidth and 
$\approx$\,819\,GcMAC/s of effective processing power
(as described in \S\ref{ssec:cmput}).

The number of frequency channels
which could be processed on a given node
was primarily determined by 
the availability of PCIe connections for data transfer.
With the then-current Intel C612 chipset,
the number of PCIe lanes available
corresponded to each node hosting four GPUs,
each of which processed a single frequency channel.

\begin{figure}[ht]
    \centering
    \includegraphics[width=\textwidth]{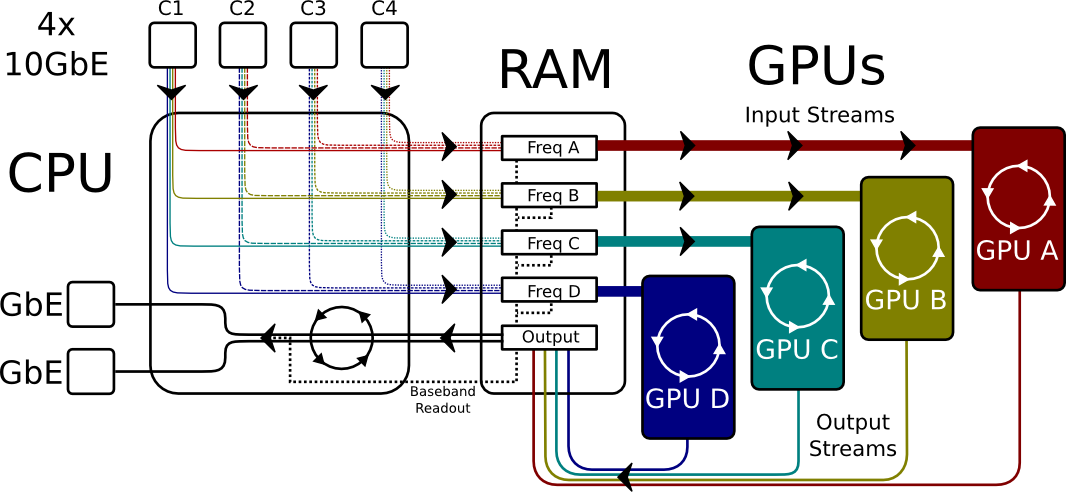}
    \caption{
    Schematic overview of data flow through a single CHIME GPU node.
    Channelized data enters on 4\,$\times$\,10\,GbE links, each from one cylinder
    and containing four frequency bands.
    These are pre-processed and assembled into full-array single-frequency sets,
    each of which is then dispatched to one of four GPUs for correlation.
    Post-correlation-and-integration data is then
    assembled and exported over standard gigabit Ethernet.
    A separate buffer of the raw data is maintained in RAM
    for triggered readout in the event of an FRB detection.
    }
    \label{fig:node-overview}
\end{figure}

\begin{figure}[ht]
    \centering
    \includegraphics[width=\textwidth]{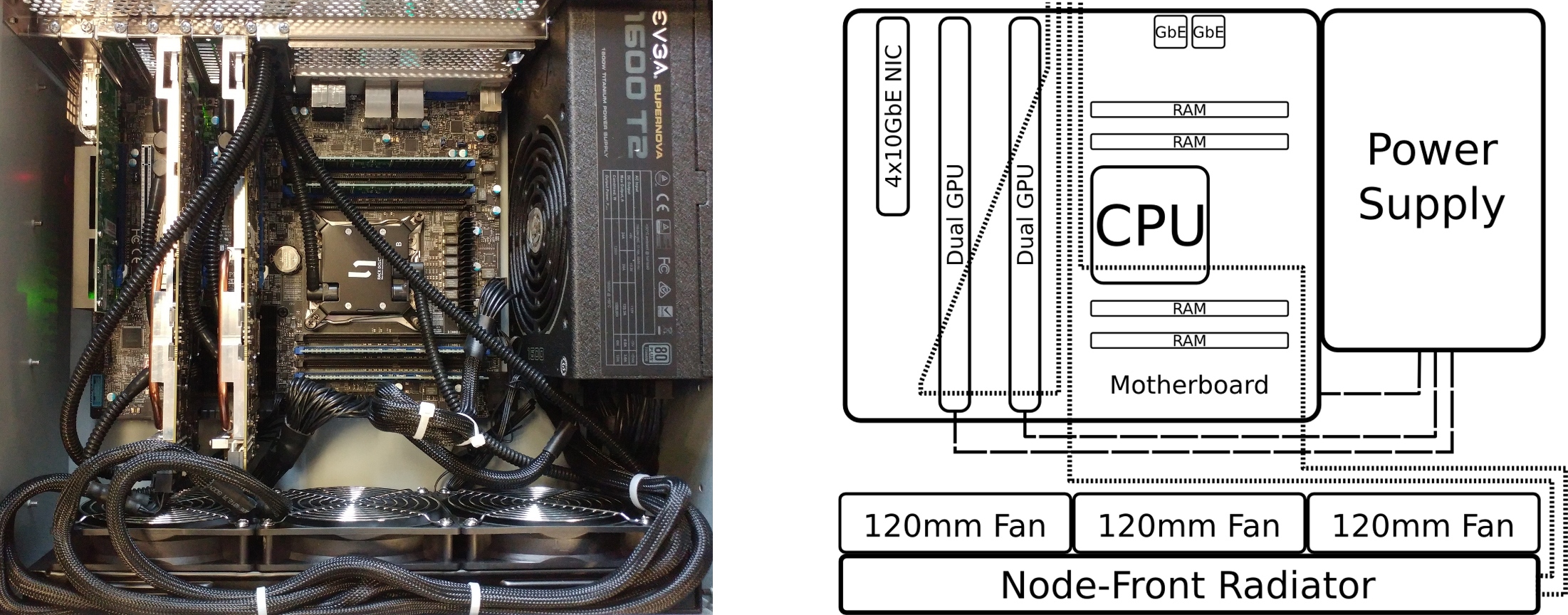}
    \caption{
        Interior of a CHIME GPU processing node (left)
        and labeled overview diagram (right)
        Power cables (black, fabric-jacketed) and coolant hoses (black, ribbed)
        are visible throughout the node, 
        and are represented in the diagram 
        by long- and short-dashed lines, respectively.
    }
    \label{fig:node-photo}
\end{figure}

A schematic overview of the data flow within a GPU node 
is shown in Figure~\ref{fig:node-overview}:
each node gathers 4\,$\times$\,10\,GbE links from the F-engine,
each containing a set of four (not necessarily adjacent) frequencies
from one of the four cylinders.
Its CPU completes the transpose,
aligning the data according to its metadata timestamp and
sending a complete frequency band to each of the four GPUs for processing.
It then exports the integrated, processed data over a pair of GbE links.
The hardware components which were selected for the nodes are listed in Table~\ref{tab:components};
an assembled node's interior is shown in Figure~\ref{fig:node-photo}.
A description of each component, 
and the reasons for each particular choice,
appear below.

\begin{wstable}[ht]
	\centering
    \begin{tabular}{ll}
        \toprule
        Component   & Hardware Selected \\
        \midrule
        GPUs        & 2\,$\times$\,AMD FirePro S9300x2 \\
        CPU         & Intel Xeon E5-2620v3 \\
        NIC         & Silicom PE310G4i71L-XR \\
        Motherboard & Supermicro X10SRA-F \\
        Memory      & 8\,$\times$\,16GB DDR4 \\
        Power Supply& EVGA SuperNOVA 1600 T2 \\
        Chassis     & General Technics CS893 \\
        Cooling 	& CoolIT Custom-Built \\
        \bottomrule
        \end{tabular}
    \caption{
        The hardware configuration of a CHIME X-engine processing node.
    }
    \label{tab:components}
\end{wstable}

\paragraph{GPUs}
The AMD FirePro S9300x2 supports two GPU processing chips per board, 
allowing a dense configuration and supplying the required processing power 
for correlation and auxiliary processing tasks.
The specific requirements of the correlation algorithm
led to the selection of the then-current AMD `Fiji' GPU platform
as it had the performance and instructions required.
Our choice of GPU was greatly influenced by 
the availability of the `MAD24' operation,
which allowed us to 
effectively perform multiple simultaneous low-bit-depth 
arithmetic operations;
these improved the $\mathcal{O}(N^2)$ correlation efficiency by a factor of two
at the cost of $\mathcal{O}(N)$ additional bookkeeping,
a substantial performance improvement for our use case \citep{2015arXiv150306203K}.

\paragraph{CPU}
Our processing and data transportation requirements
(particularly PCIe lane requirements)
limited the choice of CPU to the then-current Intel line-up.
The selected Intel Xeon E5-2620v3
supports 40 lanes of PCIe3 traffic (required for our data transfer) 
and has 6 cores operating at 2.4\,GHz, sufficient for all the processing and data manipulation we require.
Its power consumption is extremely low (85\,W TDP), 
and it is substantially less expensive than its consumer-oriented i7 counterpart.

\paragraph{Network Interface}
We required a network interface card (NIC) to provide 4\,$\times$\,10\,GbE inputs 
with Enhanced Small Form Factor Pluggable (SFP+) physical connectors.
The Silicom PE310G4i71L-XR, a quad-10\,GbE NIC built around Intel XL710 chipset,
supports the DPDK\footnote{\url{https://dpdk.org}} kernel bypass libraries
which we employ in our networking code.

\paragraph{Motherboard}
The primary requirement for the motherboard was the number of available PCIe connectors,
arranged so that two dual-slot GPUs and a network card could be operated simultaneously. 
Additional considerations included the memory and CPU options supported
and the presence of dual GbE ports.
The Supermicro X10SRA-F met or exceeded all our criteria
and additionally supports the Internet Protocol Management Interface (IPMI),
allowing for remote management at a sub-OS level.

\paragraph{Memory}
The primary system memory is a set of
registered and ECC-enabled Kingston KVR24R17D4/16 DDR4 DIMMs.
We originally allocated the required 64\,GiB of memory
as 4\,$\times\,$16\,GiB modules rather than 8\,$\times$\,8\,GiB
-- this made very little difference to the net cost 
but allowed for a trivial upgrade to 128\,GiB
per node (8\,$\times\,$16\,GiB), completed in November of 2018.
The additional memory allows the X-engine to buffer substantially more baseband data
($\approx31\,$seconds in total)
for replay in the event of an FRB detection \citep{2018ApJ...863...48C}.

\paragraph{Power Supply}
The EVGA SuperNOVA 1600 T2 is a high-capacity ATX-form-factor power supply,
with modular cabling and sufficient power connections to 
supply the GPUs with the 2\,$\times$\,8-pin PCI power connections they each require.
In order to reduce both power consumption and waste heat generated,
the node power supplies have substantial excess capacity.
A 1600W 80plus Titanium power supply is 
rated at $\geq 96\%$ efficiency when loaded at 1kW with 208V power;
the reduction in electrical consumption significantly outweighs the additional cost
given the system's near-continuous operation.

\paragraph{Chassis}
Custom-designed for CHIME, the General Technics CS893 
supports a 3\,$\times$\,120\,mm node-front radiator and 8 PCIe devices
in a maximally-compact footprint.
The main body of the chassis has a depth of only 38\,cm, 
with a standard 4U rack-mount profile of 42.6\,$\times$\,17.5\,cm.
The use of a standard ATX motherboard and power supply,
full-height PCIe cards,
and 120\,mm cooling fans
dictated a minimum 4U height for the chassis.
Most commercially-available 4U chassis
devoted significant space to storage devices;
the GPU nodes were designed to run without persistent storage,
and so could be much more compact.

\paragraph{Cooling}
The GPU and CPU dies are cooled by custom-designed CoolIT 
direct-contact liquid cooling blocks, 
exhausting heat to the in-rack coolant loops. 
Each node has two independent coolant loops, 
one of which connects to the two dual-chip GPUs
while the other serves the CPU and radiator.
The node-front radiator supports 3\,$\times$\,120\,mm fans
and couples air to the heat transfer fluid, 
regulating the air temperature inside the enclosure.

\subsection{Supporting Infrastructure}
\label{ssec:infra}

Infrastructure requirements followed directly
from the nodes' density and composition.
Covering the 1024 frequency channels requires 256 nodes,
divided into racks of 8 or 10 nodes
which operate as independent power and cooling entities.
These are housed in two enclosures,
each with 128 nodes in 13 racks
as well as two racks for additional networking and monitoring equipment.

\subsubsection{Racks and Enclosures}

The outermost enclosures are a pair of ISO 668 1AA-size (`40-foot') intermodal shipping containers.
Each houses a commercially-purchased Raymond EMC Faraday cage
which provides 
$\gtrsim110\,$dB of RFI shielding from sub-MHz to many-GHz frequencies.
One end of the `RFI cage' houses both
the low-pass filters through which electrical power enters the RFI cage
and the half-meter-square brass bulkhead through which data lines are routed.
In order to prevent RF leakage,
the fibre-optic data lines enter and exit the Faraday cage 
through a bulkhead inset with cylindrical waveguides
with an inner diameter of 1" and a length of 12"; 
these attenuate radiation at frequencies below $\sim6$\,GHz by 190-380\,dB \citep{atten}.
The opposite wall of the RFI cage is fitted with a human-sized door
and a honeycomb-mesh ventilation window;
it further includes a half-meter-square brass bulkhead
which permits the primary coolant lines to enter and exit the RFI cage.

Within each RFI cage, 
the nodes are further grouped into racks.
The Tripp~Lite SR4POST open-frame 45U four-post racks 
are extremely shallow (56\,cm),
as permitted by the custom diskless chassis.
The nodes are mounted on General Technics RK500 slide rails
whose modest length allows nodes to be removed easily in the restricted rack-front space.
The racks are arranged along the centre of each enclosure,
facing in alternating directions in order to avoid a large-scale air pressure gradient.

\subsubsection{Power Distribution}
The power consumed by the contents of each correlator enclosure
is provided in the form of 5 separate 208\,V-3$\phi$ power cables,
which are distributed to a set of 15 outlets along the length of the enclosure.
Each rack has a vertical-mount Power Distribution Unit (PDU) 
which breaks the main supply out into a number of 208\,V single-phase outlets
suitable for powering the nodes and associated equipment.
The Raritan PX3-5547 PDU selected enables 
network-controlled monitoring and switching of outlets,
per-outlet current limiting and alarms,
and control \& logging via SSH.

\subsubsection{Cooling System}
\label{ssec:full-cool}

\begin{figure}[ht]
    \centering
    \includegraphics[width=\textwidth]{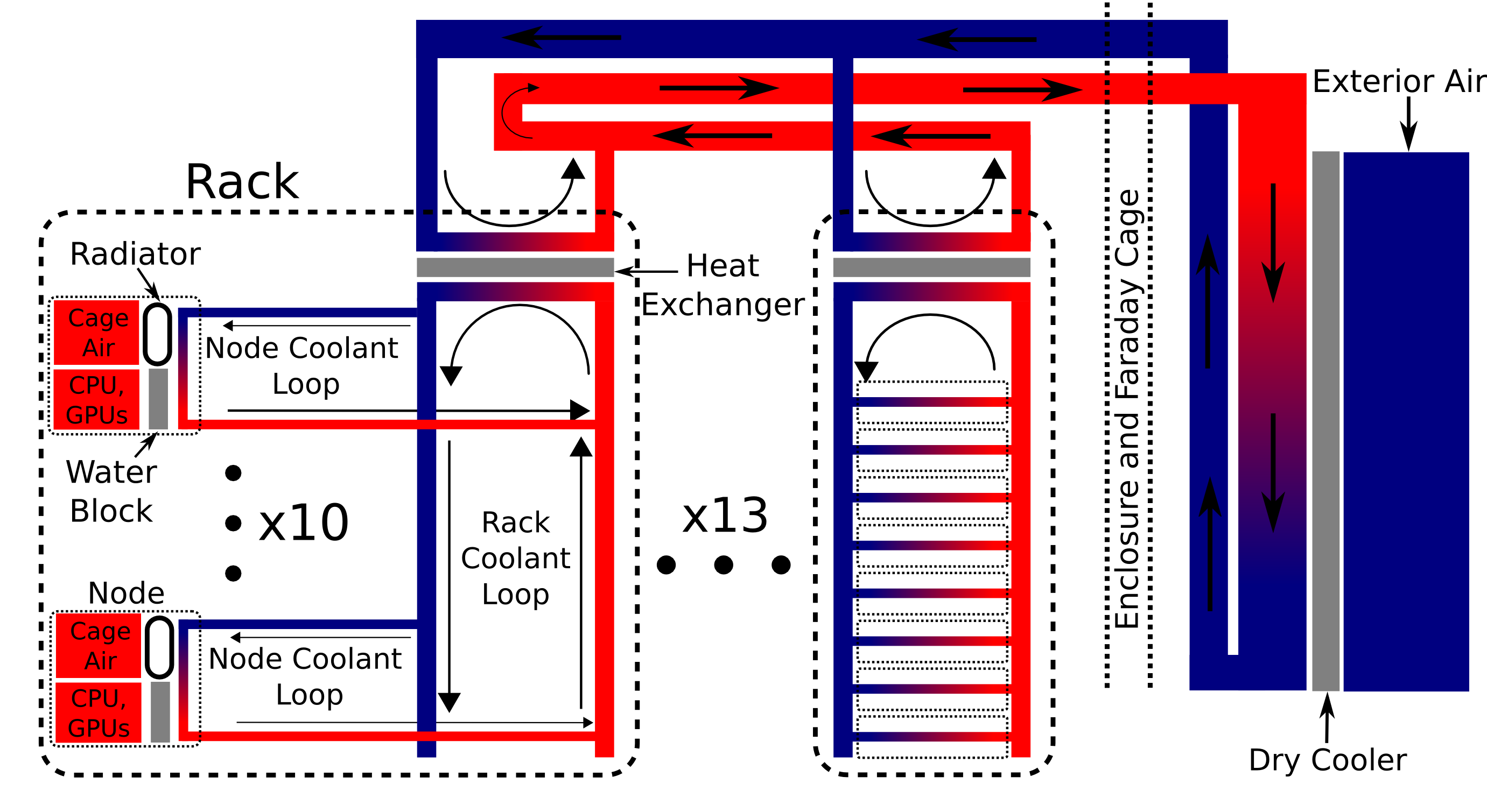}
    \caption{A diagram of the liquid-cooling system in a single enclosure,
        showing the directly-liquid-cooled components.
        The colour of a component or liquid indicates its temperature (red hot, blue cold),
        while grey objects are stages of heat transfer between parts of the system.
        The black arrows indicate the direction of fluid flow in each loop.
        }
    \label{fig:water-loop-temp}
\end{figure}

Refrigerated-air cooling options,
the traditional datacentre heat transfer solution,
require significant electrical power
and would necessitate moving large volumes of air through the enclosure's heavy RFI shielding.
The CHIME X-engine cooling system uses a liquid heat-transfer medium
to couple the heat sources inside the enclosure 
to the exterior air through a large `dry cooler'
(a Direct Coil FC07AV5D178).
A schematic of this system is shown in Figure~\ref{fig:water-loop-temp};
see \S\ref{ssec:cool-perf} for details of the system's performance.

Each rack runs an independent sealed-loop system,
which is coupled to the primary coolant through a heat exchanger.
The rack-top liquid handler (a CoolIT CHx40)
combines a heat exchanger, reservoir, \& pump
and provides basic remote monitoring and control capabilities.
The secondary coolant is distributed through a custom-engineered manifold,
and provides direct-contact liquid cooling to the CPU and GPUs 
as well as flowing through a node-front radiator 
to remove heat from the air circulating in the enclosure.

\subsubsection{Network and Management}

The input data lines connecting the F-engine and X-engine
are 256 fibre-optic `hydra' cables 
which convert the single Quad SFP+ (QSFP) connection at the F-Engine's FPGAs to
4\,$\times$\,SFP+ connectors for the GPU nodes.
Each cable is divided between four nodes 
(either within one rack
or between two adjacent racks)
so that each GPU node receives a matched set of four frequencies
from each cylinder.

The output data network is a simple hierarchical model,
with first-stage switches aggregating the GbE links from the nodes
and sending 10\,GbE links to a large central switch.
Each of the Cisco Catalyst 3650 (WS-C3650-48TQ-L) first-layer switches 
has 48\,$\times$\,GbE RJ45 and 4\,$\times$\,10\,GbE SFP+ ports;
this is sufficient to handle the 40 links from two racks' worth of nodes,
as well as individual connections to the PDUs, liquid-handler units, and file servers.
The central switch, a Cisco Nexus 3132Q (N3K-C3132Q-40GX), 
collects the 10\,GbE links from the first-stage switches
and forwards the data to dedicated servers
for final accumulation and storage.

The GPU processing nodes have no persistent storage,
instead booting and mounting filesystems 
from a set of Dell PowerEdge r410 file servers.
This simplifies software maintenance and updates,
provides a moderate level of redundancy,
and minimizes configuration overhead.
These same file servers also record logs and performance of individual nodes, 
and provide a virtual private network (VPN) for remote access.

Much of the monitoring data is supplied by
sensors built into the GPU nodes (CPU and GPU temperatures)
and PDUs (power draw), 
with additional data coming from the CHx40 units 
(coolant temperatures and flow rate, ambient air temperature and humidity).
Enclosure-wide environmental monitoring is enabled
by an NTI ENVIROMUX-5D unit
which currently connects to door-state and leak-detection sensors.

System monitoring data collection is based on the
Prometheus\footnote{\url{https://prometheus.io/}}
monitoring system,
which collects data at set intervals
and maintains a database of the time-series for each metric.
Grafana\footnote{\url{https://grafana.com/}},
a commercially-developed monitoring and data-presentation system,
is used to convert the raw sensor data in a set of 
`dashboards' summarizing the system's status.
The Prometheus system has also been configured to 
send automated alerts to 
the CHIME collaboration's internal Slack\footnote{\url{https://slack.com}} messaging system
if certain criteria are met.
Sensor values indicating severe problems
(critical overheating or coolant leaks)
automatically halt the correlator's operation,
either by stopping the main correlation software
or by directly cutting power to the affected area
via the PDUs' network interfaces.
The CPUs' temperature-driven Catastrophic Shutdown Detectors (Intel \citeyear{Intel-sdm})
and the PDU and PDC circuit breakers 
remain the automated fail-safe mechanisms of last resort.

\section{CHIME Correlator Systems Operation and Performance}
\label{sec:discuss}

\begin{figure}[ht]
	\centering
	\includegraphics[width=\textwidth]{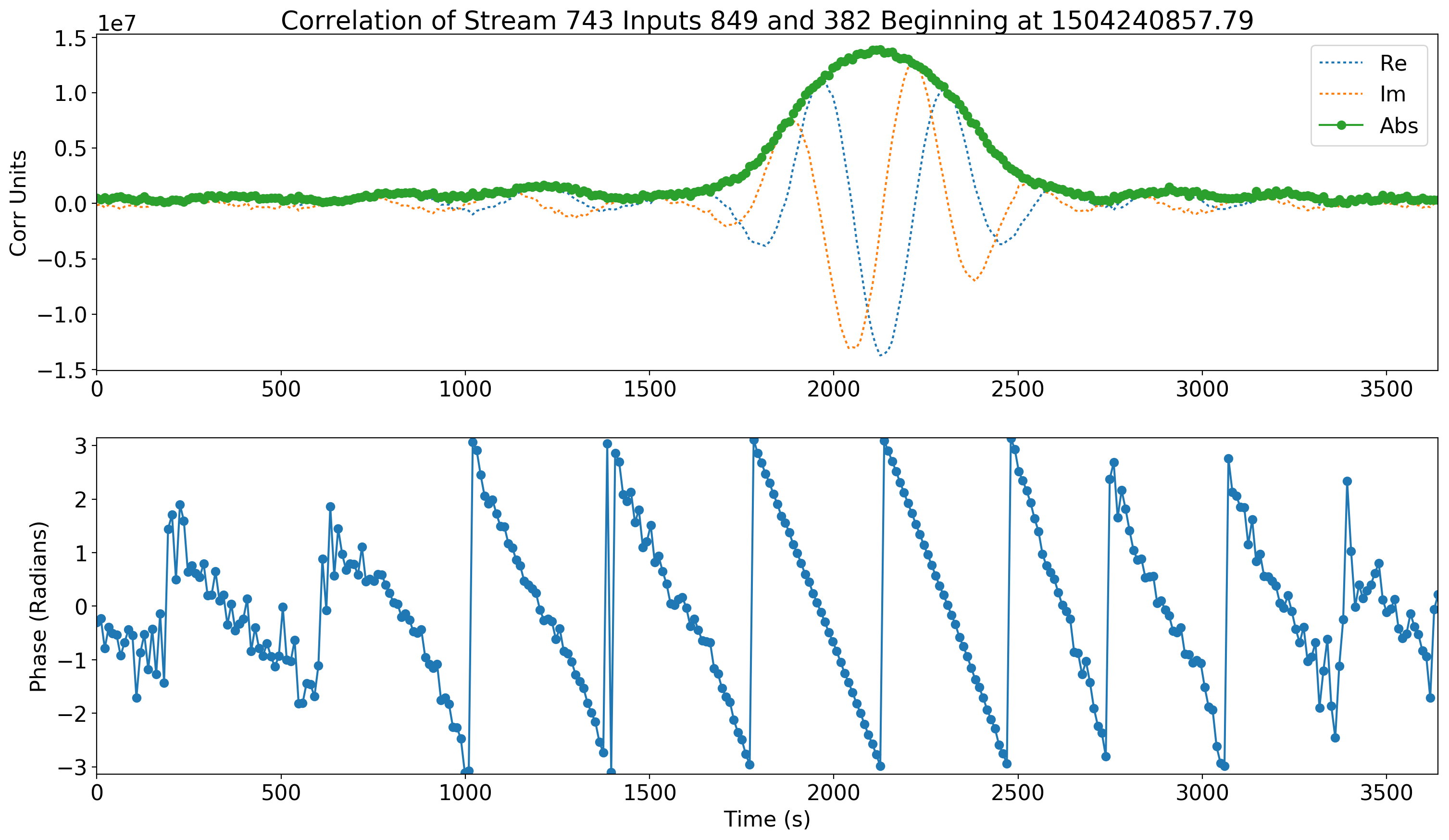}
	\caption{
	A time-stream from the `First Light' observation of Cygnus~A 
	by the full CHIME system on 31 August 2017.
	The upper panel shows the amplitude of the real and imaginary
	components of the correlation of two receivers,
	while the lower panel shows their relative phase.
	The primary beam shape produced by the cylinders is apparent in the upper panel,
	while the lower panel displays a linear phase evolution during transit
	produced by the east-west baseline separation between the cylinders.
	}
	\label{fig:first-ts}
\end{figure}

The most fundamental test of any correlator system
is if it permits the telescope to observe the sky;
the CHIME correlator emphatically succeeds at this.
An initial on-sky observation of Cygnus~A was taken on 31 August 2017.
The correlation was computed for the complete set of 2048 inputs at a subset of frequencies;
the post-outer-product data was recorded to disk in a raw, packetized state
and then transported off-site for processing and visualization.
Figure~\ref{fig:first-ts} shows a single-baseline, single-frequency timestream 
covering the entire approximately-one-hour observation.
The instrument's main beam profile
and a linear phase evolution near transit
may be readily observed;
this supported early tests
that the system was observing and correctly interpreting
an actual astronomical signal.

In addition to initial cosmological observations,
the CHIME X-engine's successful operation 
has enabled a wealth of discoveries by
the CHIME/FRB collaboration \citep{frb-n-1, frb-n-2, frb-n-3, frb-n-4};
these further validate the
end-to-end functionality of the correlator system.
A paper providing a broader overview of CHIME
and its operations
\citep{CHIME:inprep}
is forthcoming.

\subsection{Data Transfer and Processing Performance}
\label{ssec:cmput-perf}

CHIME's high-efficiency real-time data processing software,
\texttt{kotekan} \citep{Renard:inprep},
was written in OpenCL and with AMD's Heterogeneous System Architecture (HSA),
enabled by HSA compilers which translated C++ code into 
the AMD Graphics Core Next Instruction Set Architecture (GCN ISA).
The correlation task described in \S\ref{sec:x-eng-intro}
is comfortably within the available resources,
even permitting the GPUs' operation at a reduced clock rate
(722\,MHz, compared to their 975\,MHz base clock)
and reduced voltages (variable per-GPU),
which combine to vastly reduce power consumption.
Table~\ref{tab:computing} details 
the extent to which specific tasks 
contribute to the GPU's overall occupancy.


\begin{wstable}[ht]
	\centering
    \begin{tabular}{lcc}
    \toprule
    Task & Task Length (ms) & Occupancy (\%)\\
    \midrule
    Re-Order and Pre-Sum & 1.0 & 0.8\\
    RFI Excision & 1.1 & 0.9\\ 
    Correlation & 93.4 & 74.3 \\
    High-Spectral-Resolution Tap & 0.2  & 0.2\\
    Pulsar Beamforming & 5.9 & 4.7\\
    FRB Processing & 20.9 & 16.6\\
    \midrule
    Total & 122.6 & 97.4  \\
    \bottomrule
    \end{tabular}
    \caption{
    The approximate length of time required to complete 
    each processing task on the GPU,
    and the corresponding percentage of the GPU processing duty cycle it occupies.
    The tasks are measured per `block' of 49152 samples,
    $\approx125.8$\,ms in length. 
        The values presented are at a clock speed of 722\,MHz;
        this reduction from the base clock of 975\,MHz
        increases the GPUs' occupancy         
      	but consumes significantly less electrical power
      	and generates correspondingly less waste heat.
      	Additional details of the GPU nodes' 
      	software and data processing
      	may be found in
      	\citet{Renard:inprep}.
     }
    \label{tab:computing}
\end{wstable}

Table~\ref{tab:data-flow} provides typical data transfer rates
into and out of the GPU nodes.
The networking software employs DPDK,
a kernel bypass network subsystem,
to handle the extremely high rate of incoming data.
Tests show that receiving the data from the FPGAs, 
completing the transpose,
and dispatching it to the GPUs
is well within the capacity of the selected CPU;
it occupies two of the CPU's six physical cores.
Post-correlation data may be exported at a range of cadences,
but the substantial time-integration involved
ensures that the output is well within
the capacity of the two onboard GbE links.

\begin{wstable}[ht]
	\centering
    \begin{tabular}{llcc}
    \toprule
    \multicolumn{2}{c}{Data Transfer Rates} & Per Node (Gb/s) & Total (Gb/s) \\
    \midrule
    In  & Fourier-Transformed Voltage Samples & 25.6 & 6553 \\
    	& Flags and Metadata & 4.0 & 1030 \\\midrule
    Total & & 29.6 & 7583 \\\midrule\midrule
    
    Out & Correlated Visibilities (at 10\,s integration) & 0.05 & 13.6 \\ 
    	& FRB Beams & 0.59 & 153 \\
    	& Pulsar Beams & 0.25 & 64\\
    	& Flags and Metadata & 0.03 & 8.5 \\  \midrule
    Total & & 0.92 & 239\\ 
    
    \bottomrule
    \end{tabular}
    \caption{
    	Typical data rates into and out of the CHIME GPU nodes.
    	Input data for each node is carried on $4\times10$\,GbE,
    	while output data leaves on $2\,\times$\,GbE connections.
      Additional details of the GPU nodes' 
      software and data processing
      may be found in
      \citet{Renard:inprep}.
    }
    \label{tab:data-flow}
\end{wstable}

\subsection{Power Consumption}
\label{ssec:power}
Using commissioning-epoch code and the AMD ROCm 1.9.211 driver
the power consumption of a typical GPU node 
when performing both the full correlation and typical auxiliary processing
is $\approx 700$\,W,
significantly below the 1000\,W design ceiling.
Table~\ref{tab:node-pwr} details the power consumption of the GPU correlator system,
based on PDU monitoring data recorded during commissioning tests.
The X-engine's total power consumption
(with GPUs operating at a reduced clock of 722\,MHz)
is $\approx 220$\,kW,
a significant reduction from the design maximum of 256\,kW
with an accompanying reduction in operating costs.
Under typical conditions, 
$\sim 18\%$ of the system's total power is used for cooling,
which compares favorably to the 
$\gtrsim 19\%$ of the most efficient industrial systems available\footnote{
The latter value is the theoretical minimum 
for a system with an Energy Efficiency Ratio of 15;
standard datacentre planning guides \citep{lch-sysad, apc-cool} 
suggest $40\myphen50\%$ as a typical value.}.

\begin{wstable}[ht]
	\centering
    \begin{tabular}{llcc}
    \toprule
    \multicolumn{2}{c}{Power Consumption} & Per Node (W) & Total (kW) \\
    \midrule
    GPU Nodes: & GPUs & 4$\times$140 & \\
    	& CPU & 70\\
    	& Misc. & $\approx 70$ \\
    	\cmidrule(lr){2-3}
    	& Total & 700 & 180 \\
    \multicolumn{2}{l}{Infrastructure} & & 4.3 \\
    \multicolumn{2}{l}{Primary Coolant Pump} & & 5.0 \\
    \multicolumn{2}{l}{Dry Cooler Fans} & & 28.0 \\
    \midrule
    Total & & & 220 \\
    \bottomrule
    \end{tabular}
    \caption{
    	Total power consumption 
    	for the GPU correlator system
    	with all 256 nodes operating simultaneously.
    	The `infrastructure' entry includes 
    	file servers, liquid-handler units, and network equipment.    	
    }
    \label{tab:node-pwr}
\end{wstable}

\subsection{Cooling System Performance}
\label{ssec:cool-perf}

At each stage of the heat transfer system described in \S\ref{ssec:full-cool} and Fig.\,\ref{fig:water-loop-temp}, 
the thermal resistance produces a temperature gap across the heat exchanger.
In steady-state operation
the temperature of the directly-liquid-cooled components
will therefore float a fixed amount above the input coolant temperature.
Similarly, the air inside the RFI cages (and supplied to the air-cooled components)
will vary with the supplied coolant's temperature.

Tests during commissioning permitted the evaluation of these differentials; 
Table~\ref{tab:deltas} provides values extracted from a full month of temperature data.
These indicate that under representative loads
the GPUs are $20\myphen30$\,$^\circ$C warmer than the external ambient temperature
and that the internal air temperature floats  $\sim 14 \pm 3$\,$^\circ$C above the same.
The former values are still far cooler than a typical air-cooled GPU or CPU,
and so are not a source of concern.
The enclosure air temperature does, however, provide one edge of the operating envelope,
as it is the only source of cooling for a number of components.

\begin{wstable}[ht]
	\centering
	\begin{tabular}{llcc}
		\toprule
		Thermal Interface & Location & $\Delta T$ ($^\circ$C) & Cumulative $\Delta T$ ($^\circ$C)\\
		\midrule
		External Air - Primary Coolant & Dry Cooler & $\lesssim 3$ & \ndat \\
		Primary Coolant - Secondary Coolant & CHx40 & $8.7 \pm 1.9$ & \ndat \\
		Secondary Coolant - Internal Air & Radiator & 2.6 $\pm$ 2.2 & 14.3 $\pm$ 2.9 \\
		Secondary Coolant - CPU   & CPU Loop &  6.1 $\pm$ 2.5 &  17.8 $\pm$ 3.1 \\
		Secondary Coolant - GPU 0 & GPU Loop &  8.7 $\pm$ 2.3 &  20.4 $\pm$ 3.0 \\
		Secondary Coolant - GPU 1 & GPU Loop & 12.2 $\pm$ 2.7 &  23.9 $\pm$ 3.3 \\
		Secondary Coolant - GPU 2 & GPU Loop & 16.1 $\pm$ 3.4 &  27.8 $\pm$ 3.9 \\
		Secondary Coolant - GPU 3 & GPU Loop & 19.2 $\pm$ 4.0 &  30.9 $\pm$ 4.4 \\
		\bottomrule
	\end{tabular}
	\caption{
		Representative temperature differentials at various stages of the heat transfer system,
		based on commissioning tests from February 2018.
		As the external air temperature was $\sim0$\,$^\circ$C during this period,
		the dry cooler was run well below full capacity;
		the maximum $\Delta T$ value listed is from the manufacturer's specifications.
		The CHx40 liquid-handler units' $\Delta T$ was found to 
		vary linearly with the racks' power consumption;
		the value in the table corresponds to the nominal 7\,kW per rack.
		The sequentially increasing temperature of the GPUs is due to
		their serial coolant flow.
	}
	\label{tab:deltas}
\end{wstable}

The external-internal air temperature differential is higher than
expected;
the manufacturer's specifications for the CHx40 liquid-handler units
were derived from flawed simulations 
which predicted a thermal gap of $0.2^\circ$C/kW 
rather than the measured $1.24 \pm 0.27\,^\circ$C/kW.
This presents an obstacle to mid-day operation
during the hottest weeks of the year,
as the internal air temperature would be high enough
to damage more thermally-sensitive server and switch components;
the system will be required to cease operation during these periods.
The cooling system is otherwise able to 
exhaust the entire system's waste heat
and maintain appropriate internal temperatures
with minimal power consumption.

\section{Conclusion}
\label{sec:conc}

The CHIME correlator X-engine is capable of 
correlating the instrument's 2048 receiver-polarizations
over a full 400\,MHz of bandwidth.
This requires 0.8\,PcMAC/s of computation on
$7.6$\,Tb/s of input from the F-engine,
which is accomplished by the system's 256 nodes
using commercially-available GPUs as
computationally powerful, cost-effective 
units for large-scale integer matrix products.
The system consumes a modest 220\,kW of electrical power,
including that used to run the fresh-air liquid-cooling system
which keeps the GPUs at temperatures far below those of
typical air-cooled solutions.

The success of the CHIME correlator X-engine
may be illustrated by the results it has enabled;
although the primary cosmological mission is ongoing \citep{2014SPIE.9145E..22B, CHIME:inprep},
the CHIME/FRB component has made epochal contributions to FRB observations.
These include
the first FRBs seen in its frequency range \citep{frb-n-2},
the second-known repeating FRB \citep{frb-n-1},
the first known periodic FRB \citep{frb-n-3},
and the localization of an FRB to a magnetar within our own galaxy \citep{frb-n-4}.

The CHIME correlator X-Engine
achieves its design requirements
with high efficiency,
modest initial cost, 
and low power consumption (for a system of this size).
It takes full advantage of
the highly-efficient integer computation 
available on modern GPUs;
commercial PCIe network cards 
provide a reliable and cost-effective
interface capable of handling the 
immense intra-correlator bandwidth required,
while
mass-market network switches
capable of handling full-rate transfer of GbE and 10\,GbE over standard Ethernet
make the output data handling quite straightforward.
Additional details of the system hardware, layout, and performance
are available in \citet{denman-thesis}.

The system's attributes are largely dictated 
by the specifications of the CHIME instrument;
the number of receivers, bandwidth, bit depth, and F-engine structure
all have non-trivial implications for the overall design.
Although derived implementations
will therefore require careful planning 
and design modifications,
similar GPU-based X-Engines
may provide 
powerful, inexpensive, and computationally efficient
correlation for future radio interferometers.

\subsection{Future Development}
\label{ssec:future}

The CHIME spatial processing system
could only make use of the technologies
available at the time of its development;
future correlators may
realize extensive improvements
by incorporating more-recently-developed technologies.
Extensions of the CHIME correlator 
to support additional capabilities and receivers are under examination, 
and lessons learned during CHIME development 
will be applied to future correlator systems.
Additionally, continued optimization of the GPU kernels
is on course to enable further reduction of the GPU core clocks,
and therefore significant reductions in power consumption.

The continuing adoption of PCIe v4.0
by hardware manufacturers and
the increasing number of PCIe lane supported on mass-market CPUs 
loosens the NIC-GPU data transfer bottleneck;
particularly for arrays with relatively larger bandwidth
but fewer receivers,
this may substantially improve the efficiency of a GPU-based X-engine design.

The introduction of `tensor cores',
GPU components optimized for low-bit-depth integer
matrix multiplication and accumulation,
offers a potential order-of-magnitude acceleration
for correlation,
and is the subject of active research and development
\citep{ptx-isa, romein-gtc-talk}.

GPU-based correlation,
particularly with the advent
of tensor cores and
improved data-transfer technologies,
offers great potential for
high-efficiency, low-cost
digital signal processing.
The offers an unprecedented opportunity
for large-scale radio interferometers
with significantly reduced cost and improved performance.

\section*{Acknowledgements}
The authors would like to acknowledge development contributions from 
Advanced Micro Devices (AMD) 
and CoolIT Systems.

\bibliography{x-eng}
\end{document}